\documentclass[preprint,prc,letterpaper,11pt,twoside,tightenlines,nofootinbib,showpacs]{revtex4-1}
\usepackage{amsmath,amssymb,bm}
\usepackage{color}
\usepackage{graphicx}

\newcommand{\dd}{\mathrm{d}}
\newcommand{\MeV}{\mathrm{MeV}}
\newcommand{\GeV}{\mathrm{GeV}}
\newcommand{\TeV}{\mathrm{TeV}}

\newcommand{\bvec}[1]{\ensuremath{\boldsymbol{#1}}}
\newcommand{\erw}[1]{\ensuremath { %
    \left \langle {#1} \right \rangle}}

\begin{document}

\preprint{}

\title{Charm quark transport in Pb+Pb reactions at $\sqrt{s}_{NN}
  =2.76\, \TeV$ from a (3+1) dimensional hybrid approach}

\author{Thomas Lang$^{1,2}$} \author{Hendrik van Hees$^{1,2}$} \author{Jan Steinheimer$^{3}$}
\author{Marcus Bleicher$^{1,2}$}
\affiliation{
  $^{1}\,$Frankfurt Institute for Advanced Studies
  (FIAS),Ruth-Moufang-Str. 1, 60438 Frankfurt am Main, Germany }
\affiliation{
  $^{2}\,$Institut f\"ur Theoretische Physik, Johann Wolfgang
  Goethe-Universit\"at, Max-von-Laue-Str. 1, 60438 Frankfurt am Main,
  Germany }
\affiliation{
  $^{3}\,$Lawrence Berkeley National Laboratory, 1 Cyclotron Road, Berkeley, CA 94720, USA
}

\date{August 8, 2012}

\begin{abstract}
  We implement a Langevin approach for the transport of charm quarks in
  the UrQMD (hydrodynamics + Boltzmann) hybrid model. Due to the
  inclusion of event-by-event fluctuations and a full (3+1) dimensional
  hydrodynamic evolution, this approach provides a more realistic model
  for the evolution of the matter produced in heavy ion collisions as
  compared to simple homogeneous fireball expansions usually employed.
  As drag and diffusion coefficients we use a resonance approach for
  elastic heavy-quark scattering and assume a decoupling temperature of
  the charm quarks from the hot medium of $130\, \MeV$. A coalescence
  approach at the decoupling temperature for the hadronization of the
  charm quarks to D-mesons is also included.  We present calculations of
  the nuclear modification factor $R_{AA}$ as well as the elliptic flow
  $v_2$ in Pb+Pb collisions at $\sqrt {s_{NN}}=2.76\, \TeV$. The
  comparison to ALICE measurements shows a very good agreement with our
  calculations.
\end{abstract}

\maketitle


\section{Introduction}
\label{sec:Langevin}
\markright{Thomas Lang, Hendrik van Hees, Jan Steinheimer, and Marcus Bleicher. Heavy quark transport in Pb+Pb at $\sqrt{s} =2.76\,$TeV}

One major goal of high-energy heavy-ion physics is to recreate the phase
of deconfined matter, where quarks and gluons more quasi free (the Quark
Gluon Plasma, QGP) as it might have existed a few microseconds after the
Big Bang. Various experimental facilities have been built to explore the
properties of this QGP experimentally, while on the theory side a
multitude of (potential) signatures and properties of the QGP have been
predicted \cite{Adams:2005dq,Adcox:2004mh,Muller:2012zq}.

Heavy quarks are an ideal probe for the QGP. They are produced in the
primordial hard collisions of the nuclear reaction and therefore probe
the created medium during its entire evolution process. When the system
cools down they hadronize, and their decay products can finally be
detected. Therefore, heavy-quark observables provide new insights into
the interaction processes within the hot and dense medium. Two of the
most interesting observables are the elliptic flow, $v_2$, and the
nuclear modification factor, $R_{AA}$, of open-heavy-flavor mesons and
their decay products like ``non-photonic'' single electrons.  The
measured large elliptic flow, $v_2$, of open-heavy-flavor mesons and the
``non-photonic single electrons or muons'' from their decay underline
that heavy quarks take part in the collective motion of the bulk medium,
consisting of light quarks and gluons. The nuclear modification factor
shows a large suppression of the open-heavy flavor particles' spectra at
high transverse momenta ($p_T$) compared to the findings in pp
collisions. This also supports a high degree of thermalization of the
heavy quarks with the bulk medium. A quantitative analysis of the degree
of thermalization of heavy-quark degrees of freedom in terms of the
underlying microscopic scattering processes thus leads to an
understanding of the mechanisms underlying the large coupling strength
of the QGP and the corresponding transport properties.

In this letter, we explore the medium modification of heavy-flavor
transverse momentum ($p_T$) spectra. In contrast to previous studies,
see
e.g. \cite{Aichelin:2012ww,Gossiaux:2012th,Gossiaux:2010yx,Uphoff:2011ad,Uphoff:2012gb,
  Moore:2004tg,Greco:2011fs,Gossiaux:2011ea,Rapp:2009my,vanHees:2008gj,Rapp:2008fv,
  Rapp:2008qc,Greco:2007sz,vanHees:2007me,vanHees:2007mf,Vitev:2007jj,Young:2011ug,He:2011yi,He:2012xz},
we perform the simulation based on a hybrid model, consisting of the
Ultra-relativistic Quantum Molecular Dynamics (UrQMD) and a (3+1)
dimensional hydrodynamical model to simulate the bulk medium. This
approach includes event-by-event initial-state fluctuations and a full
(3+1)-dimensional hydrodynamics. The heavy-quark propagation in the
medium is described within a relativistic Langevin approach.

\section{Description of the model}

The UrQMD hybrid model has been developed to combine the advantages of
transport theory and (ideal) fluid dynamics \cite{Petersen:2008dd}. It
uses initial conditions, generated by the UrQMD model
\cite{Bass:1999tu,Dumitru:1999sf}, for a full (3+1) dimensional ideal
fluid dynamical evolution, including the explicit propagation of the
baryon current. After a Cooper-Frye transition back to the transport
description, the freeze out of the system is treated dynamically within
the UrQMD approach. The hybrid model has been successfully applied to
describe particle yields and transverse dynamics from AGS to LHC
energies
\cite{Petersen:2008dd,Steinheimer:2007iy,Steinheimer:2009nn,Petersen:2010cw,Petersen:2011sb}
and is therefore a reliable model for the flowing background medium.

The equation of state we use for our calculations includes quark and
gluonic degrees of freedom coupled to a hadronic parity-doublet model
\cite{Steinheimer:2011ea}. It includes a smooth crossover at low baryon
densities between an interacting hadronic system and a quark gluon
plasma. The thermal properties of the EoS are in agreement with lattice
QCD results at vanishing baryon density, and the EoS therefore is well
suited for our investigation at LHC energies.

The diffusion of a ``heavy particle'' in a medium consisting of ``light
particles'' can be described with help of a Fokker-Planck equation
\cite{Svet88,MS97,Moore:2004tg,HR05a,HGR05a,vanHees:2007me,Gossiaux:2008jv,He:2011yi}
as an approximation of the collision term of the corresponding Boltzmann
equation. It can be mapped into an equivalent stochastic Langevin
equation, suitable for numerical simulations. In the relativistic realm
such a Langevin process reads
\begin{equation}
\begin{split}
\label{lang.1}
\dd x_j &= \frac{p_j}{E} \dd t, \\
\dd p_j &= -\Gamma p_j \dd t + \sqrt{\dd t} C_{jk} \rho_k.
\end{split}
\end{equation}
Here $E=\sqrt{m^2+\bvec{p}^2}$, and $\Gamma$ is the drag or friction
coefficient. The covariance matrix, $C_{jk}$, of the fluctuating force
is related with the diffusion coefficients. Both coefficients are
dependent on $(t,\bvec{x},\bvec{p})$ and are defined in the (local) rest
frame of the fluid. The $\rho_k$ are Gaussian-normal distributed random
variables, i.e., its distribution function reads
\begin{equation}
\label{lang.2}
P(\bvec{\rho}) = \left (\frac{1}{2 \pi} \right)^{3/2} \exp
\left(-\frac{\bvec{\rho}^2}{2} \right ).
\end{equation}
The fluctuating force thus obeys
\begin{equation}
\label{lang.3}
\erw{F_j^{(\text{fl})}(t)}=0, \quad \erw{F_j^{(\text{fl})}(t)
  F_k^{(\text{fl})}(t')} = C_{jl} C_{kl} \delta(t-t').
\end{equation}
It is important to note that with these specifications the random
process is not yet uniquely determined since one has to specify, at
which argument of the momentum the covariance matrix $C_{jk}$ has to be
taken to define the stochastic time integral in (\ref{lang.1}). Thus, we
set
\begin{equation}
\label{lang.4}
C_{jk} =C_{jk}(t,\bvec{x},\bvec{p}+\xi \dd \bvec{p}).
\end{equation}
For $\xi=0$, $\xi=1/2$, and $\xi=1$ the corresponding Langevin processes
are called the pre-point Ito, the mid-point Stratonovic-Fisk, and the
post-point Ito (or H\"anggi-Klimontovich) realization.

The drag and diffusion coefficients for the heavy-quark propagation
within this framework are taken from a resonance approach
\cite{HR05a}.

The initial production of charm quarks in our approach is based on a
Glauber approach. For the realization of the initial collision dynamics
we use the UrQMD model. We perform a first UrQMD run excluding
interactions between the colliding nuclei and save the nucleon-nucleon
collision space-time coordinates.  These coordinates are used in a
second, full UrQMD run as possible production coordinates for the charm
quarks.

The momentum distribution for the initially produced charm quarks serves
as the starting point of our calculations. The $p_T$ distribution is
obtained from a fit to PYTHIA calculations. The fitting function for
charm quarks with $2.76\,$TeV is:
\begin{equation}
\frac{\dd N}{\dd^2 p_T} =A_1 \frac{1}{(1+A_1\cdot \left(p_T^2\right)^2)^{A_3}}
\end{equation}
with the coefficients $A_1=0.136$, $A_2=\,2.055$ and $A_3=\,2.862$.
Starting with this distribution as initial condition, at each
UrQMD/hydro time-step we perform an Ito-postpoint time-step, as
described in Sec.~\ref{sec:Langevin}. We use the UrQMD/hydro's cell
velocities, cell temperature, the size of the time-step, and the
$\gamma$-factor for the calculation of the momentum transfer,
propagating all quarks independently. Our approach provides us only with
the charm-quark distribution. Since charm quarks cannot be measured
directly in experiments we include a hadronization mechanism for
D-Mesons, via the use of a quark-coalescence mechanism. To implement
this coalescence we perform our Langevin calculation until the
decoupling temperature is reached. Subsequently we add the momenta of
light quarks to those of the charm quarks. On average the velocity of
light quarks can be approximated by the flow-velocity vector of the
local hydro cell. The mass of the light quarks is assumed to be $369 \;
\MeV$ so that the D-Meson mass becomes $1.869 \; \GeV$ when adding the
masses of the light quarks and the charm quarks ($1.5 \; \GeV$).

\section{Elliptic flow $v_2$ and nuclear modification factor $R_{AA}$}

We have performed our calculations in Pb+Pb collisions at $\sqrt{s}_{NN}
=2.76\; \TeV$ in a centrality range of 30\%-50\%. The analysis is done
in a rapidity cut of $|y|<0.35$ in line with the ALICE data.

\begin{figure}[h]
\center
\includegraphics[width=0.5\textwidth]{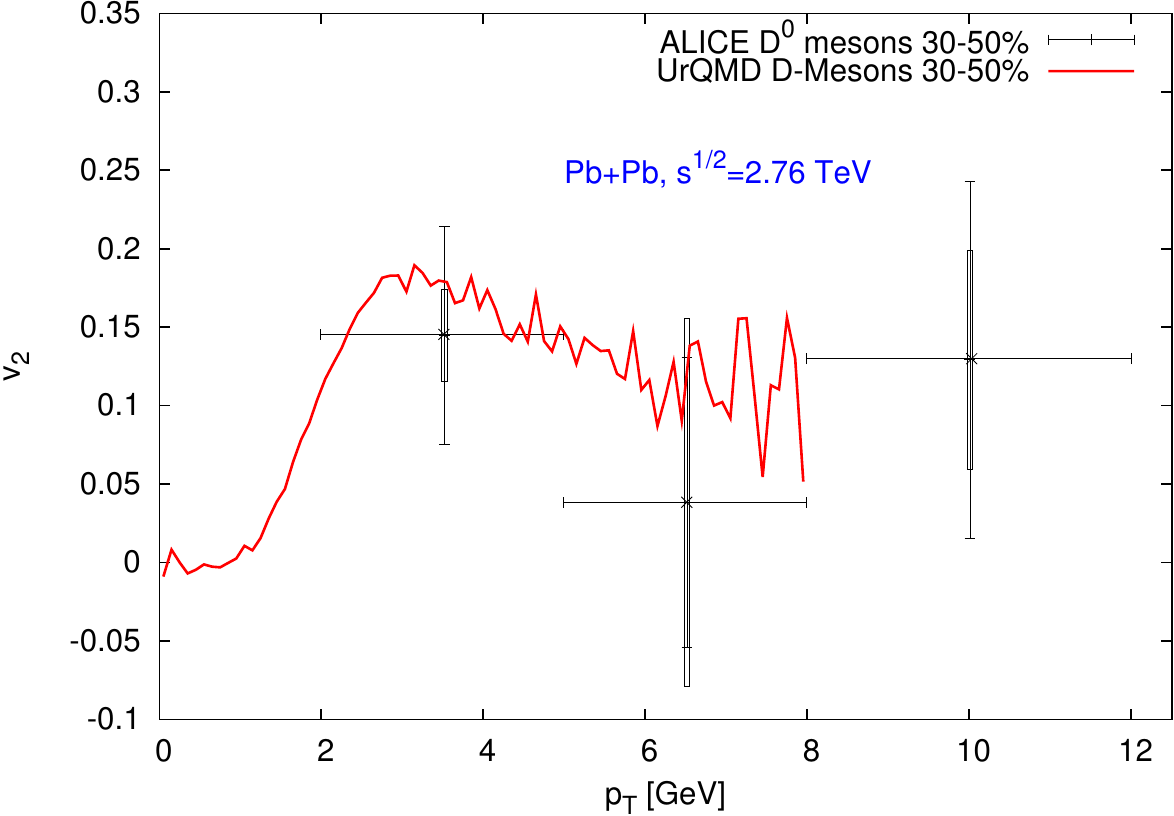}
\caption{Flow $v_2$ of D-Mesons in Pb+Pb collisions at $\sqrt
  {s_{NN}}=2.76\,$TeV compared to data from the ALICE experiment
  \cite{Bianchin:2011fa}. A rapidity cut of $|y|<0.35$ is employed.}
\label{flowLHC4}
\end{figure}

Fig.\ \ref{flowLHC4} depicts our results for the elliptic flow compared
to ALICE measurements.
The D-Meson $v_2$ exhibits a strong increase and reaches a maximum at
about $p_T=3\; \GeV$ with $v_2\sim 15\%$. Considering the error bars the
agreement between the measurements and our calculation is quite
satisfactory.

\begin{figure}[h]
\center
\includegraphics[width=0.5\textwidth]{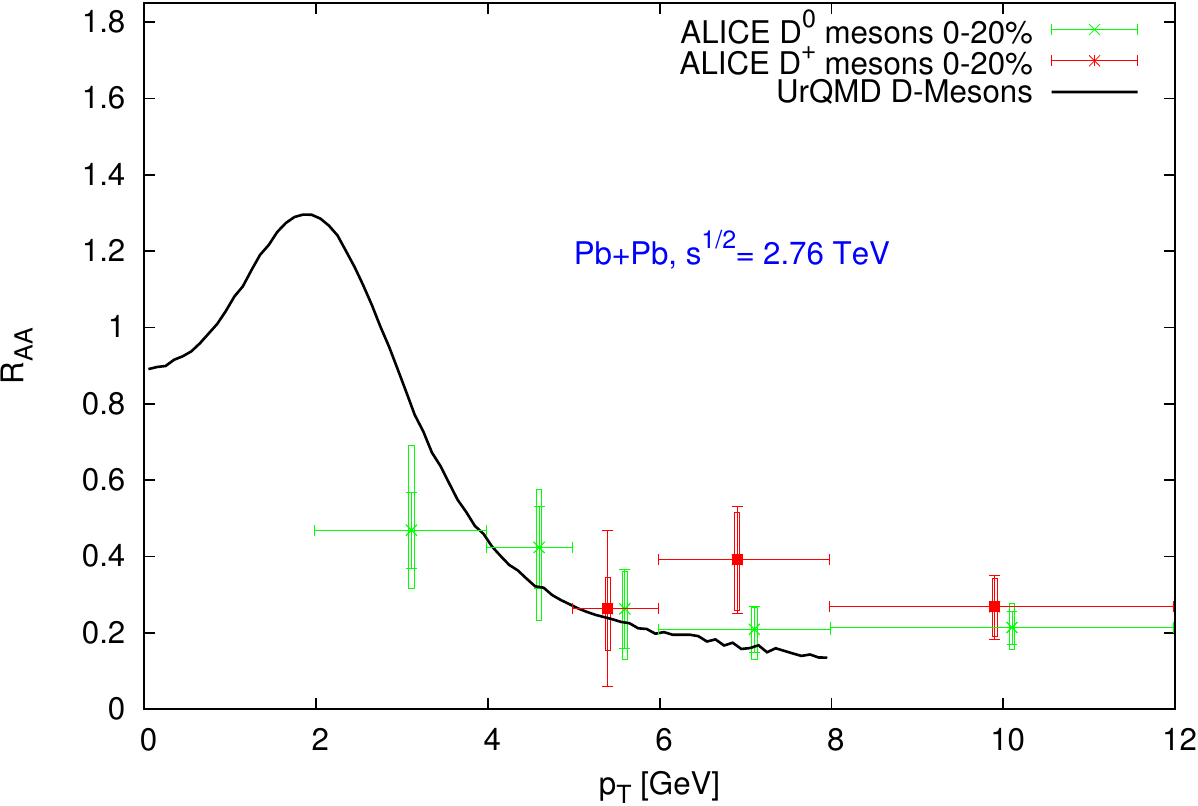}
\caption{$R_{AA}$ of D-Mesons in Pb~Pb collisions at $\sqrt
  {s_{NN}}=2.76\; \TeV$ compared to preliminary data from ALICE
  \cite{Rossi:2011nx}. A rapidity cut of $|y|<0.35$ is employed.}
\label{RAALHC4}
\end{figure}

A complementary view on the drag and diffusion coefficients is provided
by the nuclear suppression factor $R_{AA}$.  Figure \ref{RAALHC4} shows
the calculated nuclear modification factor $R_{AA}$ of D-Mesons at
LHC. Here we compare to two data sets available, for $D^0$ and $D^+$
mesons. In line with the experimental data the simulation is done for a
more central bin of $\sigma/\sigma_ {to}=0\%$-$20\%$.

We find a maximum of the $R_{AA}$ at about $2 \; \GeV$ followed by a
sharp decline to an $R_{AA}$ of about $0.2$ at high $p_T$. Especially at
low $p_T$ new measurements would be helpful to conduct a more detailed
comparison to the model prediction.

In summary we can conclude that our description of the medium
modification of charm quarks at LHC energies for both the elliptic flow
$v_2$ and the nuclear modification factor $R_{AA}$ is compatible with
the experimental measurements of ALICE.

\section{ACKNOWLEDGMENTS}

We are grateful to the Center for Scientific Computing (CSC) at
Frankfurt for providing computing resources. T.~Lang gratefully
acknowledges support from the Helmholtz Research School on Quark Matter
Studies.  This work is supported by the Hessian LOEWE initiative through
the Helmholtz International Center for FAIR (HIC for
FAIR). J.~S. acknowledges a Feodor Lynen fellowship of the Alexander von
Humboldt foundation.  This work is supported by the Office of Nuclear
Physics in the US Department of Energy's Office of Science under
Contract No. DE-AC02-05CH11231. The computational resources have been
provided by the LOEWE Frankfurt Center for Scientific Computing
(LOEWE-CSC).

%


\begin{thebibliography}{37}%
\makeatletter
\providecommand \@ifxundefined [1]{%
 \@ifx{#1\undefined}
}%
\providecommand \@ifnum [1]{%
 \ifnum #1\expandafter \@firstoftwo
 \else \expandafter \@secondoftwo
 \fi
}%
\providecommand \@ifx [1]{%
 \ifx #1\expandafter \@firstoftwo
 \else \expandafter \@secondoftwo
 \fi
}%
\providecommand \natexlab [1]{#1}%
\providecommand \enquote  [1]{``#1''}%
\providecommand \bibnamefont  [1]{#1}%
\providecommand \bibfnamefont [1]{#1}%
\providecommand \citenamefont [1]{#1}%
\providecommand \href@noop [0]{\@secondoftwo}%
\providecommand \href [0]{\begingroup \@sanitize@url \@href}%
\providecommand \@href[1]{\@@startlink{#1}\@@href}%
\providecommand \@@href[1]{\endgroup#1\@@endlink}%
\providecommand \@sanitize@url [0]{\catcode `\\12\catcode `\$12\catcode
  `\&12\catcode `\#12\catcode `\^12\catcode `\_12\catcode `\%12\relax}%
\providecommand \@@startlink[1]{}%
\providecommand \@@endlink[0]{}%
\providecommand \url  [0]{\begingroup\@sanitize@url \@url }%
\providecommand \@url [1]{\endgroup\@href {#1}{\urlprefix }}%
\providecommand \urlprefix  [0]{URL }%
\providecommand \Eprint [0]{\href }%
\providecommand \doibase [0]{http://dx.doi.org/}%
\providecommand \selectlanguage [0]{\@gobble}%
\providecommand \bibinfo  [0]{\@secondoftwo}%
\providecommand \bibfield  [0]{\@secondoftwo}%
\providecommand \translation [1]{[#1]}%
\providecommand \BibitemOpen [0]{}%
\providecommand \bibitemStop [0]{}%
\providecommand \bibitemNoStop [0]{.\EOS\space}%
\providecommand \EOS [0]{\spacefactor3000\relax}%
\providecommand \BibitemShut  [1]{\csname bibitem#1\endcsname}%
\let\auto@bib@innerbib\@empty
\bibitem [{\citenamefont {Adams}\ \emph {et~al.}()\citenamefont {Adams} \emph
  {et~al.}}]{Adams:2005dq}%
  \BibitemOpen
  \bibfield  {author} {\bibinfo {author} {\bibfnamefont {J.}~\bibnamefont
  {Adams}} \emph {et~al.} (\bibinfo {collaboration} {STAR Collaboration}),\
  }\href@noop {} {\bibfield  {journal} {\bibinfo  {journal} {Nucl. Phys. A}\
  }\textbf {\bibinfo {volume} {757}},\ \bibinfo {pages} {102}}\BibitemShut
  {NoStop}%
\bibitem [{\citenamefont {Adcox}\ \emph {et~al.}()\citenamefont {Adcox} \emph
  {et~al.}}]{Adcox:2004mh}%
  \BibitemOpen
  \bibfield  {author} {\bibinfo {author} {\bibfnamefont {K.}~\bibnamefont
  {Adcox}} \emph {et~al.} (\bibinfo {collaboration} {PHENIX Collaboration}),\
  }\href@noop {} {\bibfield  {journal} {\bibinfo  {journal} {Nucl. Phys. A}\
  }\textbf {\bibinfo {volume} {757}},\ \bibinfo {pages} {184}}\BibitemShut
  {NoStop}%
\bibitem [{\citenamefont {M{\"u}ller}\ \emph {et~al.}(2012)\citenamefont
  {M{\"u}ller}, \citenamefont {Schukraft},\ and\ \citenamefont
  {Wyslouch}}]{Muller:2012zq}%
  \BibitemOpen
  \bibfield  {author} {\bibinfo {author} {\bibfnamefont {B.}~\bibnamefont
  {M{\"u}ller}}, \bibinfo {author} {\bibfnamefont {J.}~\bibnamefont
  {Schukraft}}, \ and\ \bibinfo {author} {\bibfnamefont {B.}~\bibnamefont
  {Wyslouch}},\ }\href@noop {} {\  (\bibinfo {year} {2012})},\ \Eprint
  {http://arxiv.org/abs/1202.3233} {arXiv:1202.3233 [hep-ex]} \BibitemShut
  {NoStop}%
\bibitem [{\citenamefont {Aichelin}\ \emph {et~al.}(2012)\citenamefont
  {Aichelin}, \citenamefont {Gossiaux},\ and\ \citenamefont
  {Gousset}}]{Aichelin:2012ww}%
  \BibitemOpen
  \bibfield  {author} {\bibinfo {author} {\bibfnamefont {J.}~\bibnamefont
  {Aichelin}}, \bibinfo {author} {\bibfnamefont {P.}~\bibnamefont {Gossiaux}},
  \ and\ \bibinfo {author} {\bibfnamefont {T.}~\bibnamefont {Gousset}},\
  }\href@noop {} {\  (\bibinfo {year} {2012})},\ \Eprint
  {http://arxiv.org/abs/1201.4192} {arXiv:1201.4192 [nucl-th]} \BibitemShut
  {NoStop}%
\bibitem [{\citenamefont {Gossiaux}\ \emph {et~al.}(2012)\citenamefont
  {Gossiaux}, \citenamefont {Aichelin},\ and\ \citenamefont
  {Gousset}}]{Gossiaux:2012th}%
  \BibitemOpen
  \bibfield  {author} {\bibinfo {author} {\bibfnamefont {P.}~\bibnamefont
  {Gossiaux}}, \bibinfo {author} {\bibfnamefont {J.}~\bibnamefont {Aichelin}},
  \ and\ \bibinfo {author} {\bibfnamefont {T.}~\bibnamefont {Gousset}},\
  }\href@noop {} {\bibfield  {journal} {\bibinfo  {journal}
  {Prog.Theor.Phys.Suppl.}\ }\textbf {\bibinfo {volume} {193}},\ \bibinfo
  {pages} {110} (\bibinfo {year} {2012})}\BibitemShut {NoStop}%
\bibitem [{\citenamefont {Gossiaux}\ \emph {et~al.}(2010)\citenamefont
  {Gossiaux}, \citenamefont {Aichelin}, \citenamefont {Gousset},\ and\
  \citenamefont {Guiho}}]{Gossiaux:2010yx}%
  \BibitemOpen
  \bibfield  {author} {\bibinfo {author} {\bibfnamefont {P.}~\bibnamefont
  {Gossiaux}}, \bibinfo {author} {\bibfnamefont {J.}~\bibnamefont {Aichelin}},
  \bibinfo {author} {\bibfnamefont {T.}~\bibnamefont {Gousset}}, \ and\
  \bibinfo {author} {\bibfnamefont {V.}~\bibnamefont {Guiho}},\ }\href@noop {}
  {\bibfield  {journal} {\bibinfo  {journal} {J. Phys. G}\ }\textbf {\bibinfo
  {volume} {37}},\ \bibinfo {pages} {094019} (\bibinfo {year}
  {2010})}\BibitemShut {NoStop}%
\bibitem [{\citenamefont {Uphoff}\ \emph {et~al.}(2011)\citenamefont {Uphoff},
  \citenamefont {Fochler}, \citenamefont {Xu},\ and\ \citenamefont
  {Greiner}}]{Uphoff:2011ad}%
  \BibitemOpen
  \bibfield  {author} {\bibinfo {author} {\bibfnamefont {J.}~\bibnamefont
  {Uphoff}}, \bibinfo {author} {\bibfnamefont {O.}~\bibnamefont {Fochler}},
  \bibinfo {author} {\bibfnamefont {Z.}~\bibnamefont {Xu}}, \ and\ \bibinfo
  {author} {\bibfnamefont {C.}~\bibnamefont {Greiner}},\ }\href@noop {}
  {\bibfield  {journal} {\bibinfo  {journal} {Phys. Rev. C}\ }\textbf {\bibinfo
  {volume} {84}},\ \bibinfo {pages} {024908} (\bibinfo {year}
  {2011})}\BibitemShut {NoStop}%
\bibitem [{\citenamefont {Uphoff}\ \emph {et~al.}(2012)\citenamefont {Uphoff},
  \citenamefont {Fochler}, \citenamefont {Xu},\ and\ \citenamefont
  {Greiner}}]{Uphoff:2012gb}%
  \BibitemOpen
  \bibfield  {author} {\bibinfo {author} {\bibfnamefont {J.}~\bibnamefont
  {Uphoff}}, \bibinfo {author} {\bibfnamefont {O.}~\bibnamefont {Fochler}},
  \bibinfo {author} {\bibfnamefont {Z.}~\bibnamefont {Xu}}, \ and\ \bibinfo
  {author} {\bibfnamefont {C.}~\bibnamefont {Greiner}},\ }\href@noop {} {\
  (\bibinfo {year} {2012})},\ \Eprint {http://arxiv.org/abs/1205.4945}
  {arXiv:1205.4945 [hep-ph]} \BibitemShut {NoStop}%
\bibitem [{\citenamefont {Moore}\ and\ \citenamefont
  {Teaney}(2005)}]{Moore:2004tg}%
  \BibitemOpen
  \bibfield  {author} {\bibinfo {author} {\bibfnamefont {G.~D.}\ \bibnamefont
  {Moore}}\ and\ \bibinfo {author} {\bibfnamefont {D.}~\bibnamefont {Teaney}},\
  }\href@noop {} {\bibfield  {journal} {\bibinfo  {journal} {Phys. Rev. C}\
  }\textbf {\bibinfo {volume} {71}},\ \bibinfo {pages} {064904} (\bibinfo
  {year} {2005})}\BibitemShut {NoStop}%
\bibitem [{\citenamefont {Greco}\ \emph {et~al.}(2012)\citenamefont {Greco},
  \citenamefont {van Hees},\ and\ \citenamefont {Rapp}}]{Greco:2011fs}%
  \BibitemOpen
  \bibfield  {author} {\bibinfo {author} {\bibfnamefont {V.}~\bibnamefont
  {Greco}}, \bibinfo {author} {\bibfnamefont {H.}~\bibnamefont {van Hees}}, \
  and\ \bibinfo {author} {\bibfnamefont {R.}~\bibnamefont {Rapp}},\ }\href@noop
  {} {\bibfield  {journal} {\bibinfo  {journal} {AIP Conf.Proc.}\ }\textbf
  {\bibinfo {volume} {1422}},\ \bibinfo {pages} {117} (\bibinfo {year}
  {2012})}\BibitemShut {NoStop}%
\bibitem [{\citenamefont {Gossiaux}\ \emph {et~al.}(2011)\citenamefont
  {Gossiaux}, \citenamefont {Vogel}, \citenamefont {van Hees}, \citenamefont
  {Aichelin}, \citenamefont {Rapp} \emph {et~al.}}]{Gossiaux:2011ea}%
  \BibitemOpen
  \bibfield  {author} {\bibinfo {author} {\bibfnamefont {P.~B.}\ \bibnamefont
  {Gossiaux}}, \bibinfo {author} {\bibfnamefont {S.}~\bibnamefont {Vogel}},
  \bibinfo {author} {\bibfnamefont {H.}~\bibnamefont {van Hees}}, \bibinfo
  {author} {\bibfnamefont {J.}~\bibnamefont {Aichelin}}, \bibinfo {author}
  {\bibfnamefont {R.}~\bibnamefont {Rapp}},  \emph {et~al.},\ }\href@noop {} {\
   (\bibinfo {year} {2011})},\ \Eprint {http://arxiv.org/abs/arXiv:1102.1114
  [hep-ph]} {arXiv:1102.1114 [hep-ph]} \BibitemShut {NoStop}%
\bibitem [{\citenamefont {Rapp}\ and\ \citenamefont {van
  Hees}(2009)}]{Rapp:2009my}%
  \BibitemOpen
  \bibfield  {author} {\bibinfo {author} {\bibfnamefont {R.}~\bibnamefont
  {Rapp}}\ and\ \bibinfo {author} {\bibfnamefont {H.}~\bibnamefont {van
  Hees}},\ }\href@noop {} {\  (\bibinfo {year} {2009})},\ \bibinfo {note}
  {published in R. C. Hwa, X.-N. Wang (Ed.), Quark Gluon Plasma 4, World
  Scientific, p. 111},\ \Eprint {http://arxiv.org/abs/arXiv:0903.1096 [hep-ph]}
  {arXiv:0903.1096 [hep-ph]} \BibitemShut {NoStop}%
\bibitem [{\citenamefont {van Hees}\ \emph {et~al.}(2009)\citenamefont {van
  Hees}, \citenamefont {Mannarelli}, \citenamefont {Greco},\ and\ \citenamefont
  {Rapp}}]{vanHees:2008gj}%
  \BibitemOpen
  \bibfield  {author} {\bibinfo {author} {\bibfnamefont {H.}~\bibnamefont {van
  Hees}}, \bibinfo {author} {\bibfnamefont {M.}~\bibnamefont {Mannarelli}},
  \bibinfo {author} {\bibfnamefont {V.}~\bibnamefont {Greco}}, \ and\ \bibinfo
  {author} {\bibfnamefont {R.}~\bibnamefont {Rapp}},\ }\href@noop {} {\bibfield
   {journal} {\bibinfo  {journal} {Eur. Phys. J.}\ }\textbf {\bibinfo {volume}
  {61}},\ \bibinfo {pages} {799} (\bibinfo {year} {2009})}\BibitemShut
  {NoStop}%
\bibitem [{\citenamefont {Rapp}\ \emph {et~al.}(2008)\citenamefont {Rapp},
  \citenamefont {Cabrera}, \citenamefont {Greco}, \citenamefont {Mannarelli},\
  and\ \citenamefont {van Hees}}]{Rapp:2008fv}%
  \BibitemOpen
  \bibfield  {author} {\bibinfo {author} {\bibfnamefont {R.}~\bibnamefont
  {Rapp}}, \bibinfo {author} {\bibfnamefont {D.}~\bibnamefont {Cabrera}},
  \bibinfo {author} {\bibfnamefont {V.}~\bibnamefont {Greco}}, \bibinfo
  {author} {\bibfnamefont {M.}~\bibnamefont {Mannarelli}}, \ and\ \bibinfo
  {author} {\bibfnamefont {H.}~\bibnamefont {van Hees}},\ }\href@noop {} {\
  (\bibinfo {year} {2008})},\ \Eprint {http://arxiv.org/abs/0806.3341}
  {arXiv:0806.3341 [hep-ph]} \BibitemShut {NoStop}%
\bibitem [{\citenamefont {Rapp}\ and\ \citenamefont {van
  Hees}(2008)}]{Rapp:2008qc}%
  \BibitemOpen
  \bibfield  {author} {\bibinfo {author} {\bibfnamefont {R.}~\bibnamefont
  {Rapp}}\ and\ \bibinfo {author} {\bibfnamefont {H.}~\bibnamefont {van
  Hees}},\ }\href@noop {} {\  (\bibinfo {year} {2008})},\ \bibinfo {note}
  {published in The Physics of Quarks: New Research, Nova Publishers (Horizons
  in World Physics, Vol. 265) (2009)},\ \Eprint
  {http://arxiv.org/abs/0803.0901} {arXiv:0803.0901 [hep-ph]} \BibitemShut
  {NoStop}%
\bibitem [{\citenamefont {Greco}\ \emph {et~al.}(2007)\citenamefont {Greco},
  \citenamefont {van Hees},\ and\ \citenamefont {Rapp}}]{Greco:2007sz}%
  \BibitemOpen
  \bibfield  {author} {\bibinfo {author} {\bibfnamefont {V.}~\bibnamefont
  {Greco}}, \bibinfo {author} {\bibfnamefont {H.}~\bibnamefont {van Hees}}, \
  and\ \bibinfo {author} {\bibfnamefont {R.}~\bibnamefont {Rapp}},\ }\href@noop
  {} {\  (\bibinfo {year} {2007})},\ \Eprint {http://arxiv.org/abs/0709.4452}
  {arXiv:0709.4452 [hep-ph]} \BibitemShut {NoStop}%
\bibitem [{\citenamefont {van Hees}\ \emph {et~al.}(2008)\citenamefont {van
  Hees}, \citenamefont {Mannarelli}, \citenamefont {Greco},\ and\ \citenamefont
  {Rapp}}]{vanHees:2007me}%
  \BibitemOpen
  \bibfield  {author} {\bibinfo {author} {\bibfnamefont {H.}~\bibnamefont {van
  Hees}}, \bibinfo {author} {\bibfnamefont {M.}~\bibnamefont {Mannarelli}},
  \bibinfo {author} {\bibfnamefont {V.}~\bibnamefont {Greco}}, \ and\ \bibinfo
  {author} {\bibfnamefont {R.}~\bibnamefont {Rapp}},\ }\href@noop {} {\bibfield
   {journal} {\bibinfo  {journal} {Phys. Rev. Lett.}\ }\textbf {\bibinfo
  {volume} {100}},\ \bibinfo {pages} {192301} (\bibinfo {year}
  {2008})}\BibitemShut {NoStop}%
\bibitem [{\citenamefont {van Hees}\ \emph {et~al.}(2007)\citenamefont {van
  Hees}, \citenamefont {Greco},\ and\ \citenamefont {Rapp}}]{vanHees:2007mf}%
  \BibitemOpen
  \bibfield  {author} {\bibinfo {author} {\bibfnamefont {H.}~\bibnamefont {van
  Hees}}, \bibinfo {author} {\bibfnamefont {V.}~\bibnamefont {Greco}}, \ and\
  \bibinfo {author} {\bibfnamefont {R.}~\bibnamefont {Rapp}},\ }\href@noop {}
  {\  (\bibinfo {year} {2007})},\ \Eprint {http://arxiv.org/abs/0706.4456}
  {arXiv:0706.4456 [hep-ph]} \BibitemShut {NoStop}%
\bibitem [{\citenamefont {Vitev}\ \emph {et~al.}(2007)\citenamefont {Vitev},
  \citenamefont {Adil},\ and\ \citenamefont {van Hees}}]{Vitev:2007jj}%
  \BibitemOpen
  \bibfield  {author} {\bibinfo {author} {\bibfnamefont {I.}~\bibnamefont
  {Vitev}}, \bibinfo {author} {\bibfnamefont {A.}~\bibnamefont {Adil}}, \ and\
  \bibinfo {author} {\bibfnamefont {H.}~\bibnamefont {van Hees}},\ }\href@noop
  {} {\bibfield  {journal} {\bibinfo  {journal} {J. Phys. G}\ }\textbf
  {\bibinfo {volume} {34}},\ \bibinfo {pages} {S769} (\bibinfo {year}
  {2007})}\BibitemShut {NoStop}%
\bibitem [{\citenamefont {Young}\ \emph {et~al.}(2011)\citenamefont {Young},
  \citenamefont {Schenke}, \citenamefont {Jeon},\ and\ \citenamefont
  {Gale}}]{Young:2011ug}%
  \BibitemOpen
  \bibfield  {author} {\bibinfo {author} {\bibfnamefont {C.}~\bibnamefont
  {Young}}, \bibinfo {author} {\bibfnamefont {B.}~\bibnamefont {Schenke}},
  \bibinfo {author} {\bibfnamefont {S.}~\bibnamefont {Jeon}}, \ and\ \bibinfo
  {author} {\bibfnamefont {C.}~\bibnamefont {Gale}},\ }\href@noop {} {\
  (\bibinfo {year} {2011})},\ \Eprint {http://arxiv.org/abs/1111.0647}
  {arXiv:1111.0647 [nucl-th]} \BibitemShut {NoStop}%
\bibitem [{\citenamefont {He}\ \emph {et~al.}(2011)\citenamefont {He},
  \citenamefont {Fries},\ and\ \citenamefont {Rapp}}]{He:2011yi}%
  \BibitemOpen
  \bibfield  {author} {\bibinfo {author} {\bibfnamefont {M.}~\bibnamefont
  {He}}, \bibinfo {author} {\bibfnamefont {R.~J.}\ \bibnamefont {Fries}}, \
  and\ \bibinfo {author} {\bibfnamefont {R.}~\bibnamefont {Rapp}},\ }\href@noop
  {} {\bibfield  {journal} {\bibinfo  {journal} {Phys. Lett. B}\ }\textbf
  {\bibinfo {volume} {701}},\ \bibinfo {pages} {445} (\bibinfo {year}
  {2011})}\BibitemShut {NoStop}%
\bibitem [{\citenamefont {He}\ \emph {et~al.}(2012)\citenamefont {He},
  \citenamefont {Fries},\ and\ \citenamefont {Rapp}}]{He:2012xz}%
  \BibitemOpen
  \bibfield  {author} {\bibinfo {author} {\bibfnamefont {M.}~\bibnamefont
  {He}}, \bibinfo {author} {\bibfnamefont {R.~J.}\ \bibnamefont {Fries}}, \
  and\ \bibinfo {author} {\bibfnamefont {R.}~\bibnamefont {Rapp}},\ }\href@noop
  {} {\  (\bibinfo {year} {2012})},\ \Eprint {http://arxiv.org/abs/1208.0256}
  {arXiv:1208.0256 [nucl-th]} \BibitemShut {NoStop}%
\bibitem [{\citenamefont {Petersen}\ \emph {et~al.}(2008)\citenamefont
  {Petersen}, \citenamefont {Steinheimer}, \citenamefont {Burau}, \citenamefont
  {Bleicher},\ and\ \citenamefont {St{\"o}cker}}]{Petersen:2008dd}%
  \BibitemOpen
  \bibfield  {author} {\bibinfo {author} {\bibfnamefont {H.}~\bibnamefont
  {Petersen}}, \bibinfo {author} {\bibfnamefont {J.}~\bibnamefont
  {Steinheimer}}, \bibinfo {author} {\bibfnamefont {G.}~\bibnamefont {Burau}},
  \bibinfo {author} {\bibfnamefont {M.}~\bibnamefont {Bleicher}}, \ and\
  \bibinfo {author} {\bibfnamefont {H.}~\bibnamefont {St{\"o}cker}},\
  }\href@noop {} {\bibfield  {journal} {\bibinfo  {journal} {Phys. Rev. C}\
  }\textbf {\bibinfo {volume} {78}},\ \bibinfo {pages} {044901} (\bibinfo
  {year} {2008})}\BibitemShut {NoStop}%
\bibitem [{\citenamefont {Bass}\ \emph {et~al.}(1999)\citenamefont {Bass},
  \citenamefont {Dumitru}, \citenamefont {Bleicher}, \citenamefont {Bravina},
  \citenamefont {Zabrodin} \emph {et~al.}}]{Bass:1999tu}%
  \BibitemOpen
  \bibfield  {author} {\bibinfo {author} {\bibfnamefont {S.}~\bibnamefont
  {Bass}}, \bibinfo {author} {\bibfnamefont {A.}~\bibnamefont {Dumitru}},
  \bibinfo {author} {\bibfnamefont {M.}~\bibnamefont {Bleicher}}, \bibinfo
  {author} {\bibfnamefont {L.}~\bibnamefont {Bravina}}, \bibinfo {author}
  {\bibfnamefont {E.}~\bibnamefont {Zabrodin}},  \emph {et~al.},\ }\href@noop
  {} {\bibfield  {journal} {\bibinfo  {journal} {Phys. Rev. C}\ }\textbf
  {\bibinfo {volume} {60}},\ \bibinfo {pages} {021902} (\bibinfo {year}
  {1999})}\BibitemShut {NoStop}%
\bibitem [{\citenamefont {Dumitru}\ \emph {et~al.}(1999)\citenamefont
  {Dumitru}, \citenamefont {Bass}, \citenamefont {Bleicher}, \citenamefont
  {Stoecker},\ and\ \citenamefont {Greiner}}]{Dumitru:1999sf}%
  \BibitemOpen
  \bibfield  {author} {\bibinfo {author} {\bibfnamefont {A.}~\bibnamefont
  {Dumitru}}, \bibinfo {author} {\bibfnamefont {S.}~\bibnamefont {Bass}},
  \bibinfo {author} {\bibfnamefont {M.}~\bibnamefont {Bleicher}}, \bibinfo
  {author} {\bibfnamefont {H.}~\bibnamefont {Stoecker}}, \ and\ \bibinfo
  {author} {\bibfnamefont {W.}~\bibnamefont {Greiner}},\ }\href@noop {}
  {\bibfield  {journal} {\bibinfo  {journal} {Phys. Lett. B}\ }\textbf
  {\bibinfo {volume} {460}},\ \bibinfo {pages} {411} (\bibinfo {year}
  {1999})}\BibitemShut {NoStop}%
\bibitem [{\citenamefont {Steinheimer}\ \emph {et~al.}(2008)\citenamefont
  {Steinheimer}, \citenamefont {Bleicher}, \citenamefont {Petersen},
  \citenamefont {Schramm}, \citenamefont {St{\"o}cker} \emph
  {et~al.}}]{Steinheimer:2007iy}%
  \BibitemOpen
  \bibfield  {author} {\bibinfo {author} {\bibfnamefont {J.}~\bibnamefont
  {Steinheimer}}, \bibinfo {author} {\bibfnamefont {M.}~\bibnamefont
  {Bleicher}}, \bibinfo {author} {\bibfnamefont {H.}~\bibnamefont {Petersen}},
  \bibinfo {author} {\bibfnamefont {S.}~\bibnamefont {Schramm}}, \bibinfo
  {author} {\bibfnamefont {H.}~\bibnamefont {St{\"o}cker}},  \emph {et~al.},\
  }\href@noop {} {\bibfield  {journal} {\bibinfo  {journal} {Phys. Rev. C}\
  }\textbf {\bibinfo {volume} {77}},\ \bibinfo {pages} {034901} (\bibinfo
  {year} {2008})}\BibitemShut {NoStop}%
\bibitem [{\citenamefont {Steinheimer}\ \emph {et~al.}(2010)\citenamefont
  {Steinheimer}, \citenamefont {Dexheimer}, \citenamefont {Petersen},
  \citenamefont {Bleicher}, \citenamefont {Schramm} \emph
  {et~al.}}]{Steinheimer:2009nn}%
  \BibitemOpen
  \bibfield  {author} {\bibinfo {author} {\bibfnamefont {J.}~\bibnamefont
  {Steinheimer}}, \bibinfo {author} {\bibfnamefont {V.}~\bibnamefont
  {Dexheimer}}, \bibinfo {author} {\bibfnamefont {H.}~\bibnamefont {Petersen}},
  \bibinfo {author} {\bibfnamefont {M.}~\bibnamefont {Bleicher}}, \bibinfo
  {author} {\bibfnamefont {S.}~\bibnamefont {Schramm}},  \emph {et~al.},\
  }\href@noop {} {\bibfield  {journal} {\bibinfo  {journal} {Phys. Rev. C}\
  }\textbf {\bibinfo {volume} {81}},\ \bibinfo {pages} {044913} (\bibinfo
  {year} {2010})}\BibitemShut {NoStop}%
\bibitem [{\citenamefont {Petersen}\ \emph {et~al.}(2010)\citenamefont
  {Petersen}, \citenamefont {Qin}, \citenamefont {Bass},\ and\ \citenamefont
  {M{\"u}ller}}]{Petersen:2010cw}%
  \BibitemOpen
  \bibfield  {author} {\bibinfo {author} {\bibfnamefont {H.}~\bibnamefont
  {Petersen}}, \bibinfo {author} {\bibfnamefont {G.-Y.}\ \bibnamefont {Qin}},
  \bibinfo {author} {\bibfnamefont {S.~A.}\ \bibnamefont {Bass}}, \ and\
  \bibinfo {author} {\bibfnamefont {B.}~\bibnamefont {M{\"u}ller}},\
  }\href@noop {} {\bibfield  {journal} {\bibinfo  {journal} {Phys. Rev. C}\
  }\textbf {\bibinfo {volume} {82}},\ \bibinfo {pages} {041901} (\bibinfo
  {year} {2010})}\BibitemShut {NoStop}%
\bibitem [{\citenamefont {Petersen}(2011)}]{Petersen:2011sb}%
  \BibitemOpen
  \bibfield  {author} {\bibinfo {author} {\bibfnamefont {H.}~\bibnamefont
  {Petersen}},\ }\href@noop {} {\bibfield  {journal} {\bibinfo  {journal}
  {Phys. Rev. C}\ }\textbf {\bibinfo {volume} {84}},\ \bibinfo {pages} {034912}
  (\bibinfo {year} {2011})}\BibitemShut {NoStop}%
\bibitem [{\citenamefont {Steinheimer}\ \emph {et~al.}(2011)\citenamefont
  {Steinheimer}, \citenamefont {Schramm},\ and\ \citenamefont
  {St{\"o}cker}}]{Steinheimer:2011ea}%
  \BibitemOpen
  \bibfield  {author} {\bibinfo {author} {\bibfnamefont {J.}~\bibnamefont
  {Steinheimer}}, \bibinfo {author} {\bibfnamefont {S.}~\bibnamefont
  {Schramm}}, \ and\ \bibinfo {author} {\bibfnamefont {H.}~\bibnamefont
  {St{\"o}cker}},\ }\href@noop {} {\bibfield  {journal} {\bibinfo  {journal}
  {Phys. Rev. C}\ }\textbf {\bibinfo {volume} {84}},\ \bibinfo {pages} {045208}
  (\bibinfo {year} {2011})}\BibitemShut {NoStop}%
\bibitem [{\citenamefont {Svetitsky}(1988)}]{Svet88}%
  \BibitemOpen
  \bibfield  {author} {\bibinfo {author} {\bibfnamefont {B.}~\bibnamefont
  {Svetitsky}},\ }\href@noop {} {\bibfield  {journal} {\bibinfo  {journal}
  {Phys. Rev. D}\ }\textbf {\bibinfo {volume} {37}},\ \bibinfo {pages} {2484}
  (\bibinfo {year} {1988})}\BibitemShut {NoStop}%
\bibitem [{\citenamefont {Mustafa}\ \emph {et~al.}(1998)\citenamefont
  {Mustafa}, \citenamefont {Pal},\ and\ \citenamefont {Srivastava}}]{MS97}%
  \BibitemOpen
  \bibfield  {author} {\bibinfo {author} {\bibfnamefont {M.~G.}\ \bibnamefont
  {Mustafa}}, \bibinfo {author} {\bibfnamefont {D.}~\bibnamefont {Pal}}, \ and\
  \bibinfo {author} {\bibfnamefont {D.~K.}\ \bibnamefont {Srivastava}},\
  }\href@noop {} {\bibfield  {journal} {\bibinfo  {journal} {Phys. Rev. C}\
  }\textbf {\bibinfo {volume} {57}},\ \bibinfo {pages} {889} (\bibinfo {year}
  {1998})}\BibitemShut {NoStop}%
\bibitem [{\citenamefont {van Hees}\ and\ \citenamefont {Rapp}(2005)}]{HR05a}%
  \BibitemOpen
  \bibfield  {author} {\bibinfo {author} {\bibfnamefont {H.}~\bibnamefont {van
  Hees}}\ and\ \bibinfo {author} {\bibfnamefont {R.}~\bibnamefont {Rapp}},\
  }\href@noop {} {\bibfield  {journal} {\bibinfo  {journal} {Phys. Rev. C}\
  }\textbf {\bibinfo {volume} {71}},\ \bibinfo {pages} {034907} (\bibinfo
  {year} {2005})}\BibitemShut {NoStop}%
\bibitem [{\citenamefont {van Hees}\ \emph {et~al.}(2006)\citenamefont {van
  Hees}, \citenamefont {Greco},\ and\ \citenamefont {Rapp}}]{HGR05a}%
  \BibitemOpen
  \bibfield  {author} {\bibinfo {author} {\bibfnamefont {H.}~\bibnamefont {van
  Hees}}, \bibinfo {author} {\bibfnamefont {V.}~\bibnamefont {Greco}}, \ and\
  \bibinfo {author} {\bibfnamefont {R.}~\bibnamefont {Rapp}},\ }\href@noop {}
  {\bibfield  {journal} {\bibinfo  {journal} {Phys. Rev. C}\ }\textbf {\bibinfo
  {volume} {73}},\ \bibinfo {pages} {034913} (\bibinfo {year}
  {2006})}\BibitemShut {NoStop}%
\bibitem [{\citenamefont {Gossiaux}\ and\ \citenamefont
  {Aichelin}(2008)}]{Gossiaux:2008jv}%
  \BibitemOpen
  \bibfield  {author} {\bibinfo {author} {\bibfnamefont {P.~B.}\ \bibnamefont
  {Gossiaux}}\ and\ \bibinfo {author} {\bibfnamefont {J.}~\bibnamefont
  {Aichelin}},\ }\href@noop {} {\bibfield  {journal} {\bibinfo  {journal}
  {Phys. Rev. C}\ }\textbf {\bibinfo {volume} {78}},\ \bibinfo {pages} {014904}
  (\bibinfo {year} {2008})}\BibitemShut {NoStop}%
\bibitem [{\citenamefont {Bianchin}(2011)}]{Bianchin:2011fa}%
  \BibitemOpen
  \bibfield  {author} {\bibinfo {author} {\bibfnamefont {C.}~\bibnamefont
  {Bianchin}} (\bibinfo {collaboration} {ALICE Collaboration}),\ }\href@noop {}
  {\  (\bibinfo {year} {2011})},\ \Eprint {http://arxiv.org/abs/arXiv:1111.6886
  [hep-ex]} {arXiv:1111.6886 [hep-ex]} \BibitemShut {NoStop}%
\bibitem [{\citenamefont {Rossi}(2011)}]{Rossi:2011nx}%
  \BibitemOpen
  \bibfield  {author} {\bibinfo {author} {\bibfnamefont {A.}~\bibnamefont
  {Rossi}},\ }\href@noop {} {\bibfield  {journal} {\bibinfo  {journal} {J.\
  Phys.\ G}\ }\textbf {\bibinfo {volume} {38}},\ \bibinfo {pages} {124139}
  (\bibinfo {year} {2011})}\BibitemShut {NoStop}%
\end{thebibliography}

\end{document}